\documentclass[12pt]{iopart}
\expandafter\let\csname equation*\endcsname\relax
\expandafter\let\csname endequation*\endcsname\relax
\usepackage{graphicx,amsmath,algorithm,algorithmicx,algpseudocode,color}
\usepackage{subfigure,float}

\algdef{SE}[DOWHILE]{Do}{doWhile}{\algorithmicdo}[1]{\algorithmicwhile\ #1}%
\begin{document}
\title[]{Quantum Approximate Optimization Algorithm in
Non-Markovian Quantum Systems}

\author{Bo Yue, Shibei Xue}
\address{Department of Automation, Shanghai Jiao Tong University and the Key Laboratory of System Control and Information Processing, Ministry of Education of China, Shanghai, 200240, P. R. China}
\ead{yuebo2017JD@sjtu.edu.cn, shbxue@sjtu.edu.cn}
%\vspace{10pt}
\author{Yu Pan}
\address{State Key Laboratory of Industrial Control Technology, Institute of Cyber-Systems and Control,
College of Control Science and Engineering, Zhejiang University, Hangzhou 310027, P. R. China}
\ead{ypan@zju.edu.cn}
\author{Min Jiang}
\address{School of Electronics and Information Engineering, Soochow University, Suzhou 215006, P. R. China}
\ead{jiangmin08@suda.edu.cn}

\begin{abstract}
Although quantum approximate optimization algorithm (QAOA) has demonstrated its quantum supremacy, its performance on Noisy Intermediate-Scale Quantum (NISQ) devices would be influenced by complicated noises, e.g., quantum colored noises. To evaluate the performance of QAOA under these noises, this paper presents a framework for running
QAOA on non-Markovian quantum systems which are represented by an augmented system model.
In this model, a non-Markovian environment carrying quantum colored noises is modelled as an ancillary system driven by quantum
white noises which is directly coupled to the corresponding principal system; i.e., the computational unit for the algorithm.
With this model, we mathematically formulate QAOA as piecewise Hamiltonian control of the augmented system, where we also optimize the control depth to %reduce the influence of noises on the algorithm.
%\textcolor{red}{
fit into the circuit depth of current quantum devices.
%} 
For efficient simulation of QAOA in non-Markovian quantum systems, a boosted algorithm using quantum trajectory is further presented. Finally, we show that non-Markovianity can %\textcolor{red}{
be utilized as a quantum resource
%} 
to achieve a %\textcolor{red}{
relatively good performance of QAOA, which is characterized by our proposed exploration rate.
\end{abstract}

\section{\label{sec:level1}Introduction}

%量子计算重要性(5-6行）

Quantum computing has been a rapidly-growing technology since %\textcolor{red}{
Shor's algorithm for exponential speedup of large integer factorization~\cite{365700,shor1999polynomial} and Grover's algorithm for quadratic speedup of unstructured database searching~\cite{grover}
%} 
, compared to their classical counterparts.
%\textcolor{red}{
Recent works involve reduced computational time for chemistry~\cite{b3,b4,b5}, machine learning~\cite{b-6,b-7}, finance~\cite{b-10,b-11} and other fields~\cite{b-8,b-9}. However, these quantum algorithms rely on scalable, fault-tolerant, universal quantum computers, which are not currently available.
%}

%\textcolor{red}{
Current quantum devices are not advanced enough for fault-tolerance and consist of moderate number of noisy qubits. And thus current stage of QC is referred to as Noisy Intermediate-Scale Quantum (NISQ)~\cite{PhysRevA.106.010101} era.
%}
%Although these quantum algorithms work for some specific problems, it has a long way to accelerate classical algorithms for a class of combinatorial optimization problems which are  NP-complete or even NP-hard.
In recent years, a class of quantum approximation optimization algorithms (QAOAs%\textcolor{red}{s}
)~\cite{farhi2014quantum}  %\textcolor{red}{
are proposed for solving combinatorial optimization problems, which adopt a hybrid quantum-classical paradigm making use of a parameterized quantum circuit and a classical variational loop~\cite{PhysRevA.106.010101}. Instead of an optimal solution, QAOA can give
an approximated solution in a short time. Initially, QAOA ran on closed quantum systems, which was first applied to the Max-Cut problem~\cite{farhi2014quantum}, and to Max E3LIN2~\cite{Farhi2014AQA} soon after. Many advanced QAOAs have been developed, since Ref.~\cite{farhi2016quantum} pointed out that QAOA can achieve so-called quantum supremacy.
Different from the above standard QAOA employing classical variational optimizers, Ref.~\cite{magann2021feedback} proposed a feedback-based strategy to minimizing an optimization objective which has achieved fewer iterations than those in the standard one.
Further, unconstrained QAOA was extended to solve combinatorial optimization problems with soft and hard constraints~\cite{hadfield2017quantum,hadfield2019quantum}.
To evaluate the performance of QAOA, a benchmark has been established in Ref.~\cite{willsch2020benchmarking}, %\textcolor{red}{
where three different measures, i.e., the probability of finding the ground state, the energy expectation value, and a ratio closely related to the approximation ratio on weighted Max-Cut problems and 2-satisfiability problems. Also, many applications of QAOA have been found. For example, QAOA was applied to generate non-trivial quantum states~\cite{ho1803efficient, ho2018ultrafast}, to solve
the heterogenous vehicle routing problem (HVRP)~\cite{fitzek2021applying}, the lattice protein folding problem~\cite{Fingerhuth2018AQA}, as well as a clustering problem in unsupervised machine learning~\cite{otterbach2017unsupervised}.
On the other hand, QAOA has been experimentally tested on small-scale quantum processors which can be considered to be noise-free. In Ref.~\cite{bengtsson2020improved}, QAOA successfully ran on two superconducting transmon qubits for the exact-cover problem. Also, in a two qubits system,
Ref.~\cite{abrams2019implementation} found that XY interactions can help to efficiently solve specific problems by reducing the depth of QAOA. Shortly afterwards, QAOA has run on a planar superconducting processor with twenty-three qubits for solving non-planar graph problems. Although QAOA has been studied in various systems and problems, it presumes that the algorithm runs on a closed quantum system%\textcolor{red}{
~\cite{wang2018quantum,Larkin2022EvaluationOQ,10.1007/s11128-021-03342-3} and has not taken noise effects into account. However, when the problem to be solved runs on
Noisy Intermediate-Scale Quantum (NISQ) devices~\cite{Preskill2018quantumcomputingin,PhysRevA.106.010101}, the performance of the above QAOAs would degrade since quantum information carriers are inevitably affected by decoherence.

%\textcolor{red}{
Further, how to achieve a high-level performance of QAOA on NISQ devices becomes a research focus in recent years. To study this, it is important to take the model of quantum noises into account.
%} 
One feasible way is to consider open quantum systems. 
On one hand, 
%\textcolor{red}{
researchers hope to reduce the depth of QAOA and the duration of quantum evolution when the computing units are under noises. Ref.~\cite{PanXue} proposes an improved QAOA for Markovian quantum systems where control depth was minimized so as to reduce the influence of Markovian noises on the performance of the algorithm. 
% Ref.~\cite{venuti2021optimal} presents an argument for compilation policies to exploit the unique characteristics of QAOA-circuits alongside the variation-awareness of the noisy devices, which significantly improves the circuit success probability. 
On the other hand, researchers hope to analyze characteristics of noises, among which certain settings may %\textcolor{red}{
benefit the performance of quantum computing and further QAOA.%}
%\textcolor{red}{
Instead of destroying the quantum effects  that leads to the power of quantum computing, Ref.~\cite{qr1} shows dissipation can be a fully-fledged resource for universal quantum computing when driving the system to a steady state where the outcome of the computation is encoded and when engineering various strongly correlated states in steady state. Ref.~\cite{qr2} shows that the system's decay rate can be reduced, by adding generalized-Markovian noise on top of a background Markovian dynamics. In Ref.~\cite{yang2017optimizing}, optimal nonadiabatic bang-bang protocols outperform conventional quantum annealing in the presence of weak white additive external noises where the system is described by Redfield master equation. Ref.~\cite{assist} presents a noise‑assisted variational quantum thermalization that can find several systems for which the thermal state can be approximated with a high fidelity and is applicable in QAOA. 

Although the above works have taken Markovian noises into account for quantum computing or QAOA, computational units would be affected by more complicated noises, e.g., quantum colored noises, where the units can exhibit totally different dynamics%\textcolor{red}{
~\cite{xue2020inverse,tan2020quantum,8537939}.
%For system identification of quantum colored noise,
%identifies the Hamiltonian of the quantum system under colored measurement noise; Ref.~\cite{xue2020inverse,8537939,xue2017non2} identify the characteristics of the non-Markovian quantum systems influenced by quantum colored noise; Ref.~\cite{xue2017non1} generalizes the Hamiltonian identification problem to open quantum systems via gradient algorithm. 
Similar to aforementioned Markovian systems, non-Markovian evolution can be utilized as a quantum resource to boost certain indicators or the performance of the system.
Ref.~\cite{qr3} demonstrates that non-local memory effects can be effectively used to decrease the error rate of a quantum channel. This indicates that systems undergoing a non-Markovian evolution may also serve as a quantum resource to facilitate the performance of QAOA. In fact, non-Markovian quantum systems have been identified in solid-state quantum systems which should be an important class of NISQ devices, so it is necessary to study whether %\textcolor{red}{
high performance of QAOA can be achieved in non-Markovian quantum systems, where we utilize non-Markovianity as a quantum resource. However, this is still an open problem.

In this paper, we investigate QAOA in non-Markovian quantum systems.
 To capture the dynamics of a non-Markovian quantum system with an arbitrary spectrum of quantum colored noises, we model the non-Markovian quantum system by an augmented system where an ancillary system represents the internal modes of quantum colored noises. The spectrum of the fictitious output for the ancillary system is  consistent with that of a non-Markovian environment. Also, the ancillary system is coupled to the computational units (a principal system %\textcolor{red}{
 responsible for executing QAOA%}
 ) by a direct interaction such that the principal system undertakes non-Markovian dynamics. With this model, we propose a framework for QAOA in non-Markovian quantum systems where the optimizer balances between the performance of QAOA and the control depth. Further, a boosted algorithm using quantum trajectory is proposed to %\textcolor{red}{
 accelerate the calculation of density matrices for the high-dimensional augmented system.
Numerical simulations on the Max-Cut problem show that QAOA in non-Markovian quantum systems can outperform that in Markovian quantum systems and %\textcolor{red}{
QAOA performs better with non-Markovianity serving as a quantum resource, which is characterized by an appropriate degree of exploration rate%}
.

%an appropriate degree of non-Markovianity can improve the performance of QAOA.

%本文要做什么（结构）
This paper is structured as follows. Section \ref{sec:two} provides a brief review of QAOA in closed and Markovian open quantum system. The augmented system model for non-Markovian systems is reviewed in Section \ref{31}. QAOA in non-Markovian quantum systems with a depth constraint is proposed in Section \ref{32}.  Boosted QAOA is given in Section~\ref{34}. Numerical simulations on the Max-Cut problem that test the performance of our algorithm are presented in Section \ref{41} and \ref{42}, where parameter analyses for the degree of non-Markovianity and comparison of the performance of QAOA between Markovian and non-Markovian quantum systems are conducted in Section \ref{43}. In Section \ref{44} and \ref{45}, more complicated Max-Cut cases and a case with different noise are studied for generality of our algorithm. Section \ref{sec:six} concludes our work %\textcolor{red}{
and offers future research directions.%}

\section{Brief Review of QAOA in Quantum Systems: from Closed to Open}\label{sec:two}
QAOA is a hybrid quantum-classical variational algorithm designed to tackle combinatorial optimization problems~\cite{zhou2020quantum}. In this section, we briefly review noise-free (standard) QAOA in closed quantum systems, as well as noisy QAOA in Markovian quantum systems. We will first introduce how to convert a combinatorial optimization problem to a QAOA formulation and then give some basic notations in the algorithm.
\subsection{Combinatorial Optimization Problems and Ising Formulation}
Combinatorial optimization is a kind of problems that searches solutions in a discrete but huge space, which maximizes or minimizes an objective function. Typical examples entail travelling salesman problems, knapsack problems, and Max-Cut problems, etc. To deal with these problems, exhaustive search is not tractable, and only specialized algorithms which require strong techniques and approximate algorithms are feasible at current situation.

To solve combinatorial optimization problems, Ising formulation is presented in Ref.~\cite{Lucas2014IsingFO}, which can convert a classical objective function
\begin{equation}
    \mathcal{V}(x_1,x_2,\cdots,x_N)=-\sum_{i<j}J_{ij}x_ix_j+\sum_{i=1}^{N}h_ix_i,
    \label{eq:Isi}
\end{equation}
into a quantum version Hamiltonian
\begin{equation}
    H=\mathcal{V}(\sigma_1^z,\sigma_2^z,\cdots,\sigma_N^z)=\sum_{\alpha=1}^M H_\alpha.
    \label{eq:Ham}
\end{equation}
This objective (\ref{eq:Isi}) is general which can represent a large number of combinatiorial optimization problems. %\textcolor{red}{
In Eq.~(\ref{eq:Isi}), the objective involves an $N$-sites spin chain, where each spin is represented by a discrete variable $x_i \in \{-1,+1\},~i=1,\cdots,N$. Additionally, the parameter $J_{ij}$ is the coupling strength (or interaction) between two adjacent sites, namely the $i$-th and the $j$-th site, and $h_i$ denotes the local external magnetic field for the $i$-th site. Since these discrete variables take values identical to the eigenvalues of a spin system, we can replace $x_i$ with the Pauli-z matrix $\sigma_i^z$ and tensor the $2\times 2$ identity matrix $I$ for other spins.
For instance, when $N=4$, a term $-J_{24}x_2x_4$ is converted into $-J_{24}I\otimes \sigma_2^z \otimes I \otimes \sigma_4^z$, where $N$ is the total number of qubits in the algorithm and is equal to the logarithm  of the number of solutions. In a specific problem, we can further write the Hamiltonian as the summation of local cost functions $H_\alpha$ with the total number $M$. The expression of $H_\alpha$ depends on the problem. In this way, the task of minimizing the classical objective is converted to find the minimum eigenvalue of the Hamiltonian $H$. For more details, see Ref.~\cite{Lucas2014IsingFO}.
\subsection{Standard QAOA in Closed Quantum Systems}
By far, many quantum algorithms that can potentially demonstrate the so-called quantum advantage are based on ideally fault-tolerant quantum computers with low error rates and long coherence durations~\cite{RevModPhys.94.015004}. In other words, the inherent quantum system has to be a closed quantum system; i.e., a quantum system does not exchange information with other systems. The state of a closed quantum system can be described by a wave function $|\psi(t)\rangle$, whose time evolution obeys the Schr${\rm\ddot{o}}$dinger equation $i \frac{\partial}{\partial t}|\psi(t)\rangle = H|\psi(t)\rangle$. For simplicity, we let Plank's constant $\hbar$ to be $1$ hereafter. Alternatively, for a closed quantum system, its state can also be described by a density matrix $\rho_c(t)=|\psi(t)\rangle\langle\psi(t)|$ which satisfies the Liouville–von Neumann equation
\begin{equation}
    \dot{\rho_c}(t) = -i[H, \rho_c(t)],\label{noisefree}
\end{equation}
where $[\cdot, \cdot]$ is the commutator of operators.

The standard QAOA encodes all possible solutions to a combinatorial optimization problem in quantum states. The objective function of a combinatorial optimization problem is also converted into a Hamiltonian $H$ for a quantum system as shown in the above subsection. QAOA intends to approach the ground state of $H$, which is an approximate solution closed to the optimal one~\cite{wang2018introduction}. Mathematically, the standard QAOA solves the following optimization problem
\begin{equation}
    \min_{\tau=(\beta,\zeta)} f(\tau)=\min_{\tau=(\beta,\zeta)}\langle\psi(\tau)|H|\psi(\tau)\rangle,
\end{equation}
where $f(\tau)$ is the expectation of $H$ given a wave function $|\psi(\tau)\rangle$. Here, $\beta=(\beta_1,\cdots,\beta_P)^T$ and $\zeta=(\zeta_1,\cdots,\zeta_P)^T$ are two vectors of control durations for two unitary evolutions
\begin{equation}
    U(H, \zeta_\cdot) = e^{-i\zeta_\cdot H}=e^{-i\zeta_\cdot \sum_{\alpha=1}^MH_{\alpha}}=\prod_{\alpha=1}^{M}e^{-i\zeta_\cdot H_{\alpha}},\label{eq:piece2}
\end{equation}
\begin{equation}
    U(H', \beta_\cdot)  = e^{-i\beta_\cdot H'}=e^{-i\beta_\cdot \sum_{\theta=1}^N\sigma^x_{\theta}}=\prod_{\theta=1}^Ne^{-i\beta_\cdot \sigma^x_{\theta}},\label{eq:piece}
\end{equation}
where $P$ is the initial control depth of QAOA.
In terms of functionality, $U(H, \zeta_\cdot)$ alters the phase of potentially good quantum states, while the mixing non-commuting Hamiltonian 
\begin{equation}
H'=\sum_{\theta=1}^N\sigma^x_{\theta}
\end{equation}
enables the operator $U(H',\beta_\cdot)$ to rotate quantum states to change the probability or weight of these good quantum states in superposition states. In addition, $\zeta_\cdot$ and $\beta_\cdot$, elements of $\zeta$ and $\beta$, are variational parameters that indicate control durations of $H$ and $H'$. Note that in the Hamiltonian $H'$ the subscript $\theta$ in the Pauli-x matrix $\sigma^x_{\theta}$ indicates the order of qubits and thus we
can further express it as $\sigma^x_{\theta}=I_{1}\otimes \cdots \otimes I_{\theta-1} \otimes \sigma^x_{\theta} \otimes I_{\theta+1}\otimes \cdots \otimes I_{n}$ with $2\times 2$ identity matrices $I_\cdot$.

With the above notations, the standard QAOA can be described as follows. Initially, the state of qubits is prepared in a uniform superposition of all possible states $|\psi_0\rangle=|+\cdots+\rangle=|+\rangle_n\cdots|+\rangle_1$ with a superposition state  $|+\rangle=\frac{1}{\sqrt{2}}(|0\rangle+|1\rangle)$ where $|0\rangle=(1,0)^T$ and $|1\rangle=(0,1)^T$ are the excited state and the ground state for each qubit.
Further, the evolution of the states of the qubits satisfies the Liouville equation (\ref{noisefree}) where the Hamiltonian alternates between $H$ and $H'$ and the durations are determined by $\beta_\cdot$ and $\zeta_\cdot$, respectively. The above process can be described as
\begin{equation}
   {\rho_{c,0}}\stackrel{U(H, \zeta_1)}{\longrightarrow}\rho_{c,1}\stackrel{U(H', \beta_1)}{\longrightarrow}\rho_{c,2}\cdots\stackrel{U(H, \zeta_P)}{\longrightarrow}\rho_{{c,2P-1}}\stackrel{U(H, \beta_P)}{\longrightarrow}\rho_{{c,2P}}\label{eq:flow}
\end{equation}
with $\rho_{c,0}=|\psi_0\rangle\langle\psi_0|$, which is one iteration in QAOA. Here, we denote the state generated at the end of each iteration as $\rho_c(\tau)=\rho_{{c,2P}}$. The final state $\rho_c(\tau)$ is measured to optimize the duration $\tau$  through an optimizer within a classical computer for the next iteration. Finally, with sufficient iterations, the state of qubits approaching to its ground state encodes an approximate optimal solution to the objective function of a combinatorial optimization problem.
\subsection{QAOA in Markovian Quantum Systems}\label{sec:markov}

Since basically no quantum system is completely isolated from its environments and the quantum computers developed thus far have a limited coherence time. Hence, QAOA operating in a closed quantum system is far from reality. For NISQ devices, it is necessary to consider QAOA in open quantum systems.
An open quantum system interacts with an external environment or another quantum system, which significantly alters its dynamics and results in decoherence~\cite{breuer2007theory}.

In open quantum systems, the performance of QAOA was first investigated in a class of Markovian quantum systems where the system is disturbed by quantum white noises with a flat spectrum~\cite{PanXue}. Different from Eq.~(\ref{noisefree}),
the master equation
\begin{equation}
    \dot{\rho}_m(t) = -i[H, \rho_m(t)] + \sum_{n=1}^{R}\gamma_n\mathcal{L}^*_{L_n}(\rho_m(t)),\label{eq:markov}
\end{equation}
for the density matrix $\rho_m(t)$ of a Markovian quantum system has an additional Lindblad term%\textcolor{red}{
~\cite{lindblad1,lindblad2}
%}
which characterizes the dissipation process in this system. Here, $\gamma_n$ and $L_n$ are the damping rate %\textcolor{red}{
(also called decoherence rate or dissipation rate) and the coupling operator between each qubit and the environment for the $n$th dissipative channel out of a total number of $R$. The Lindblad superoperator is calculated as $\mathcal{L}^*_{L_n}(\rho_m(t)) = \frac{1}{2}([L_n\rho_m(t), L_n^{\dagger}]+[L_n,\rho_m(t)L_n^{\dagger}])$. For more specifications concerning the modelling of Markovian quantum systems, see references~\cite{article,noise,filtering}.

Since the state of open quantum systems are described by a density matrix instead of a wave function, the optimization target for QAOA in Markovian quantum systems can be mathematically formulated as
\begin{equation}
    \min_{\tau=(\beta,\zeta)} g(\tau)=\min_{\tau=(\beta,\zeta)}{\rm tr}(H \rho_m(\tau)).
\end{equation}

We aim to find a control duration vector $\tau$ such that the expectation of the Hamiltonian is minimized. The basic procedure for QAOA in Markovian quantum systems is similar to that in closed quantum systems. However, due to the decoherence process, the system does not keep a unitary evolution. Hence, when we calculate the evolution of the state, we should use Eq.~(\ref{eq:markov}). Also, since decoherence deteriorates the state of the system, it is expected to complete the algorithm in a short duration.

\section{QAOA in Non-Markovian Quantum Systems represented by an Augmented System Model}\label{sec:three}
Although Markovian quantum systems can capture parts of dynamics of open quantum systems, there exist other quantum systems involving complicated environments resulting in totally different dynamics from those of Markovian ones~\cite{PhysRevB.79.125317,chirolli2008decoherence,PhysRevA.86.052304,7605518}. For example, non-Markovian quantum systems are disturbed by quantum colored noise that are generated from the environments with memory effects. Generally, the shape of the spectrum $S(\omega)$ of a quantum colored noise is not flat such that the system and environment can exchange information. This kind of non-Markovian quantum systems has been found in solid-state quantum systems such as quantum dots or superconducting systems~\cite{chirolli2008decoherence}. To run QAOA in these solid-state systems, it is necessary to theoretically explore QAOA in non-Markovian quantum systems.

In this section, we first review the augmented system model for non-Markovian systems. Based upon this, we formulate QAOA in non-Markovian quantum systems and propose a regularized non-Markovian QAOA, whose control depth can be reduced. Further, to mitigate the taxing computation burden on classical processors simulating quantum processors, a boosted QAOA algorithm is proposed.

\subsection{Augmented System Model for Non-Markovian Quantum Systems}\label{31}

\begin{figure}[htbp]
\centering
\includegraphics[scale=0.6]{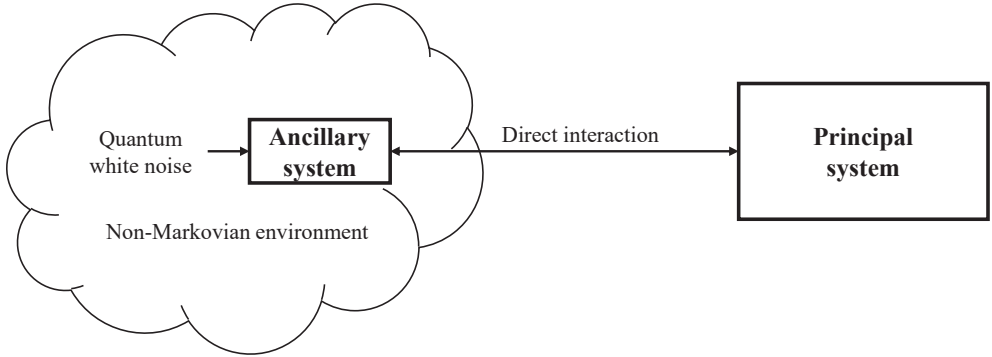}
\caption{A schematic plot of the augmented system model for non-Markovian quantum systems.}
\label{aug}
\end{figure}

As aforementioned, a non-Markovian environment is characterized by a noise spectrum $S(\omega)$ which indicates there exist internal modes in the environment. Therefore, an augmented system approach to representing a non-Markovian quantum system takes  into account the dynamics of these internal modes of the non-Markovian environment such that we can describe the dynamics of the non-Markovian system in an augmented Hilbert space~\cite{article}. A schematic plot of the augmented system model for non-Markovian quantum systems is depicted in Fig.~\ref{aug}.

%In Ref., a systematic augmented Markovian system approach is put forward to modelling non-Markovian quantum systems. This proposed non-Markovian model distinguishes itself from predecessors' non-Markovian models in that the internal quantum system and the environment mutually influences each other, and both systems interact in a direct manner instead of the classical field-mediated connection discussed in ~\cite{PhysRevA.86.043819}.
In general, the shape of the spectrum of a non-Markovian environment can be arbitrary. When the spectrum is rational, we can use quantum spectral factorization theorem to find a linear quantum system realization for the internal modes of the non-Markovian environment. Otherwise, Pade approximation method can be applied to approximate an irrational spectrum by a rational one such that we can also find its linear quantum system realization~\cite{article}. In this way, we realize a non-Markovian environment by a linear ancillary system.
The ancillary system is actually a cluster of quantum harmonic oscillators. Its Hamiltonian is $H_a=\bar{a}^\dagger\Omega_a\bar{a}$ where $\bar{a}$ is a vector of annihilation operators of the harmonic oscillators and $\Omega$ characterizes the internal energy of each oscillator as well as the interactions between them. These oscillators are driven by quantum white noise where their couplings are expressed by a vector of operators $L_a=N_a \bar{a}$ with a suitable dimensional matrix $N_a$ describing the coupling strengths. To represent the quantum colored noise generated by the non-Markovian environment, a fictitious output is introduced $c_a=K_a\bar{a}$ with a suitable dimensional matrix $K_a$. Hence, we obtain the transfer function from the quantum white noise inputs to the fictitious output as $\Gamma(s)=-K_a(sI+i\Omega_a+\frac{1}{2}N_a^\dagger N_a)^{-1}N_a^\dagger$ where $s=-i\omega$ is the complex variable in Laplace transform. Note that quantum spectral factorization theorem for determining the corresponding matrices $\omega_a$, $K_a$ and $N_a$ from a given spectrum $S(\omega)$ for the non-Markovian environment can be found in Ref.~\cite{article}.

To capture the mutual influence between a non-Markovian system and its environment, an direct interaction Hamiltonian $H_{pa} = i(c_a^\dagger z-z^\dagger c_a)$ is introduced where $z$ is a vector of operators for the principal system in the augmented system model. In such a way, the Langevin equation for the principal system can be consistent with a traditional integral-differential Langevin equation for non-Markovian quantum systems. Hence, we can represent the non-Markovian quantum system using the augmented system model~\cite{xue2017modelling}.

In the Schr${\rm\ddot{o}}$dinger picture, the augmented system model can be described by a master equation
\begin{equation}
    \dot{\rho}(t) = -i[H_p+H_a+H_{pa}, \rho(t)] +\sum_{f=1}^{Q}\gamma_{_f} \mathcal{L}^*_{ L_{a,f}}(\rho(t)),\label{eq:nonmarkov}
\end{equation}
where $\rho(t)$ is the density matrix of the augmented system defined on the tensor space of the Hilbert space for the ancillary and principal systems, $f$ denotes the order of $L_a$ when there are multiple couplings, and $\gamma_{_f}$ is the damping rate of the oscillator with respect to the $f$th dissipative channel out of a total number of $Q$. $H_p$ is the Hamiltonian for the principal system, which describes the energy of qubits in QAOA. %\textcolor{red}{
Thus, $H_p=H$, where $H$ is previously defined in Eq.~(\ref{eq:Ham}). Note that although Eq.~(\ref{eq:nonmarkov}) is written in a Markovian form, the state of the principal system obeys a non-Markovian dynamic, which can be obtained by tracing out the degree of freedom of the ancillary system from the density matrix of the augmented system model. Concretely, the density matrix of the principal system $\rho_p(t)$ can be calculated as
\begin{equation}
    \rho_p(t) = {\rm tr}_a(\rho(t))=\sum_j (I_p\otimes\langle j|_a)\rho(
    t)(I_p\otimes|j\rangle_a),\label{eq:partial}
\end{equation}
where ${\rm tr}_a$ is the partial trace with respect to the ancillary system, $I_p$ is the identity matrix of the Hilbert space of the principal system $\mathcal{H}_p$, and $\{|j\rangle_a\}$ is a set of orthogonal bases of the Hilbert space for the ancillary system $\mathcal{H}_a$.

%\textcolor{red}{
Note that although the augmented system is Markovian, the dynamics of the principal system are non-Markovian. For more details, readers can refer to Ref.~\cite{filterxue} and Ref.~\cite{xue2017modelling}.%}

\subsection{QAOA in Non-Markovian Quantum Systems}\label{32}
With the augmented system model, we can design a QAOA in non-Markovian quantum system. Similar to the standard QAOA, the objective for QAOA in non-Markovian quantum systems can be written as
 \begin{equation}
    \min_{\tau=(\beta,\zeta)} h(\tau)=\min_{\tau=(\beta,\zeta)}{\rm tr}(H\rho_p(\tau)),\label{eq:nonmark}
\end{equation}
%\textcolor{red}{
where $H$ is defined in Eq.~(\ref{eq:Ham}). Note that we should use the density matrix for the principal system $\rho_p(\tau)$ to calculate the trace of the Hamiltonian but not that for the augmented system. The density matrix for the principal system $\rho_p(\tau)$ should be obtained from Eq.~(\ref{eq:partial}). 

\begin{figure}[htbp]
    \centering
    \includegraphics[scale=0.55]{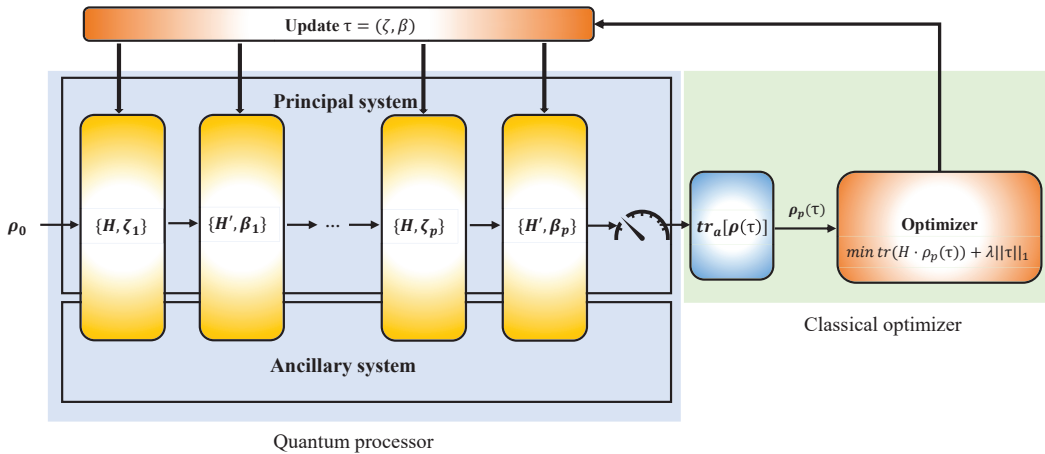}
    \caption{A framework of regularized QAOA in non-Markovian quantum systems. The purple area indicates quantum devices, while the green area indicates classical devices.}
    \label{qaoa}
\end{figure}

Although we can consider the above optimization for QAOA in non-Markovian quantum systems, it is expected to complete the calculation of QAOA in a suitable coherence time. Hence, it is required to squeeze the duration of our QAOA, so we add the $l_1$ norm of $\tau$ into the above objective function; i.e.,
\begin{equation}
    \min_{\tau=(\beta,\zeta)} y(\tau)=\min_{\tau=(\beta,\zeta)}{\rm tr}(H \rho_p(\tau))+\xi ||\tau||_1,\label{eq:all}
\end{equation}
where $||\cdot||_1$ refers to the $l_1$ norm and thus we have $||\tau||_1 = \sum_{\mu=1}^P(|\beta_\mu|+|\zeta_\mu|)$. The regularization parameter $\xi >0$ indicates a balance between the result of QAOA and its calculation duration. It can be noted that the number of control duration $P$ is not necessarily reduced since $\beta_\mu$ and $\zeta_\mu$ is not necessarily punished to zero. However, a reduction of $\beta_\mu$ and $\zeta_\mu$ is also important since cutting down control duration is beneficial to limited control capability of NISQ devices. Fig.~\ref{qaoa} shows the framework of regularized QAOA in non-Markovian quantum systems. Note that the purpose of considering depth in the objective is to complete the algorithm as fast as possible to avoid the noise effects but not to suppress the noise, which is a reasonable consideration in current quantum devices.

Traditional gradient descent method fails to find a solution to the optimization problem (\ref{eq:all}), as the $l_1$ regularization norm term is not differentiable. Hence, we adopt  a proximal gradient (PG) descent algorithm to solve the problem (\ref{eq:all}), which has been widely used in machine learning~\cite{murphy2012machine}. The major difference brought by PG descent algorithm is the soft-threshold operation, defined as
\begin{equation}
S_{\xi \upsilon}({[d_k]}_i)=\left\{
\begin{aligned}
{[d_k]}_i-\xi \upsilon & , & {[d_k]}_i>\xi \upsilon \\
0 \, \, \,\quad& , & -\xi \upsilon\leq{[d_k]}_i\leq\xi \upsilon \\
{[d_k]}_i+\xi \upsilon & , & {[d_k]}_i<-\xi \upsilon,\\
\end{aligned}
\right.
\end{equation}
where ${[d_k]}_i$ is the $i$th component of control parameters for the $k$th iteration before the soft-threshold operation.
The values of the control parameters are shifted towards zero by an amount of $\xi \upsilon$. Furthermore, the parameter will be penalized to zero, if the condition $|{[d_k]}_i|\leq\xi \upsilon$ is satisfied. By this means, less functional control parameters are eliminated so as to reduce the overall control depth.
After a number of epochs of PG descent, the minimized duration $\hat{\tau}$ and the solution to Eq.~(\ref{eq:all}), i.e., $\rho_p(\hat{\tau})$, are obtained such that the approximate minimum value and corresponding solution to the combinatorial optimization problem can be obtained.

The workflow of this regularized QAOA in non-Markovian quantum systems is described as follow. First, we randomly generate an initial duration vector $\tau_1$ and the augmented system evolves with alternate Hamiltonians with respective durations. Afterwards, the state of the principal system $\rho_p$ and the corresponding objective $h(\tau_k)$ are calculated. Here, the gradient for the $i$th segment of the duration vector $\tau$ is calculated as
\begin{equation}
    [\nabla {\rm tr}(H\rho_p(\tau))]_i=\frac{{\rm tr}(H\rho_p(\tau_i+\epsilon))-{\rm tr}(H\rho_p(\tau_i-\epsilon))}{2\epsilon},
\end{equation}\\
where $\epsilon$ is a small positive perturbation. Consequently, $\tau_2$ is updated along the gradient descent direction with a proper learning rate $v$ and followed by the soft-threshold operation. The above three steps repeat by multiple times until a terminating condition is satisfied which signifies the convergence of the objective $h(\tau)$. In the following algorithm, the terminating condition is chosen as the difference between two successive objectives and the threshold is denoted as $\eta$. The regularized QAOA in non-Markovian quantum systems is detailed in Alg.~\ref{alg:pro}, where $k$ is the current iteration step.

\begin{algorithm}[htbp]
    \caption{Regularized QAOA in non-Markovian quantum systems}%标题
    \label{alg:pro}%标签
    \begin{algorithmic}[1]
      %这里是伪代码内容
      \Require an initial duration vector $\tau_1$, a weight $\xi$, a learning rate $\upsilon$, two Hamiltonians $H$ and $H'$, a terminating threshold $\eta$
      \Ensure the updated duration vector $\tau_k$, the state of the principal system $\rho_p(\tau_k)$ and its corresponding objective value $h(\tau_k)$
      \State $k = 1$
      \State EVOLUTIONONCE($\tau_k$, $H$, $H'$)
      \Do
      \State $k \leftarrow k+1$
      \State $d_k = \tau_{k-1}-\upsilon \nabla tr(H\rho_p(\tau_{k-1}))$
      \State $\tau_{k}=S_{\xi \upsilon}(d_k)$\Comment{the soft-thresholding operation}
      \State EVOLUTIONONCE($\tau_k$, $H$, $H'$)
      \doWhile{$|h(\tau_k)-h(\tau_{k-1})|<\eta$}\Comment{the terminating condition}\\
      
      \Return $\tau_k,\rho_p(\tau_k), h(\tau_k)$\\
      \Function{Evolutiononce}{$\tau_k$, $H$, $H'$}
      \State $p=\frac{1}{2}|\tau_k|$\Comment{$|\cdot|$ denotes the vector cardinality}
      \For{$j=1,\cdots, p$}
        \State  $H_p = H$ for $\beta_{jk}$ duration in Eq.~(\ref{eq:nonmarkov}) $\rightarrow$ $\dot{\rho}$ $\rightarrow$ $\rho$
        \State $H_p = H'$ for $\zeta_{jk}$ duration in Eq.~(\ref{eq:nonmarkov}) $\rightarrow$ $\dot{\rho}$ $\rightarrow$ $\rho$
      \EndFor 
      \State $\rho_p(\tau_k)$ $=$ ${\rm tr}_a[\rho]$\Comment{trace out the ancillary system}
      \State $h(\tau_k)={\rm tr}(H\rho_p(\tau_k))$
      \State \Return $h(\tau_k)$, $\rho_p(\tau_k)$
      \EndFunction
    \end{algorithmic}
\end{algorithm}

\subsection{An Accelerated QAOA in Non-Markovian Quantum Systems Using Quantum Trajectory}\label{34}

Since the augmented system model is defined in an augmented Hilbert space $\mathcal{H}_p\otimes\mathcal{H}_a$, the dimension of the augmented system increases exponentially with the number of qubits. To avoid taxing  computation burden in calculation of the evolution of non-Markovian quantum systems for QAOA, we propose an accelerated QAOA based on quantum trajectory~\cite{k1993,daley2014}, which can save both time and memories.

To calculate the evolution of the augmented system in a quantum trajectory way, we first rewrite Eq.~(\ref{eq:nonmarkov}) in an alternative form as 
\begin{equation}
\dot{\rho}(t) =-i( H_e \rho(t)-\rho(t){H}_e^\dagger)+\sum_{f=1}^Q \gamma_{_f}L_{a,f}\rho(t)L_{a,f}^\dagger,\label{qt1}
\end{equation}
where we denote
\begin{equation}
{H}_e ={H}_p+{H}_a+{H}_{pa}-\frac{i}{2}\sum_{f=1}^Q\gamma_{_f} L_{a,f}^\dagger L_{a,f},\label{qt2}
\end{equation}
as a non-Hermitian effective Hamiltonian, defined on $\mathcal{H}_p \otimes \mathcal{H}_a$. For each individual trajectory, the initial state is $|\phi(0)\rangle=|\phi_p(0)\rangle\otimes|\phi_a(0)\rangle$ where $|\phi_p(0)\rangle$ and $|\phi_a(0)\rangle$ are the initial wave functions of the principal and ancillary systems, respectively.
%, which satisfies $\rho(0)=|\psi(0)\rangle\langle\psi(0)|$
Without loss of generality, we focus on the evolution of quantum state $|\phi(t)\rangle$ within the time step from $t$ to $t+\delta t$ in a single trajectory. $|\phi(t)\rangle$ evolves under the influence of the effective Hamiltonian $H_e$ and the jump operators $L_{a,f}$, one within each time interval. We define 
% \begin{equation}
% 1-\delta p=\langle\psi(t)|(1+i{H}_e^\dagger\delta %t)(1-i{H}_e\delta t)|\psi(t)\rangle,\label{prob}
% \end{equation}
\begin{equation}
1-\delta p=\langle\phi(t)|e^{iH_e^{*}\delta t}e^{-iH_e\delta t}|\phi(t)\rangle,\label{prob}
\end{equation}
where the probability $\delta p$ denotes the contraction in the norm of $|\phi(t)\rangle$. And thus at the time instant $t+\delta t$, the state $|\phi(t+\delta t)\rangle$ dominated by $H_e$ with a probability $1-\delta p$ is calculated as
\begin{equation}
|\phi(t+\delta t)\rangle=\frac{e^{-iH_e\delta t}|\phi(t)\rangle}{\sqrt{1-\delta p}}.\label{qt3}
\end{equation}

In addition, we rewrite Eq.~(\ref{prob}) as
\begin{equation}
    \delta p \, =  \,\langle\phi(t)|(1-e^{iH^{*}_e\delta t} e^{-iH_e\delta t})|\phi(t)\rangle
    % =&\frac{\langle\psi(t)|(L_p^\dagger L_p+L_a^\dagger L_a)|\psi(t)\rangle}{\langle\psi(t)|\psi(t)\rangle}\nonumber\\
    =  \,\delta t\langle\phi(t)|\sum_{f=1}^{Q}\gamma_{_f}L_{a,f}^\dagger L_{a,f}|\phi(t)\rangle
    \equiv  \,\sum_{f=1}^Q\delta p^{\ \downarrow}_{a,f},
\end{equation}
where $\downarrow$ denotes $\delta p_{a,f}$ are arranged in a descending order. At the time instant $t+\delta t$, the state $|\phi(t+\delta t)\rangle$ dominated by $L_{a,f}$ with a probability $\delta {p_{a,f}}$ is calculated as
\begin{equation}
|\phi(t+\delta t)\rangle=\frac{\sqrt{\gamma_{_f}}L_{a,f}|\phi(t)\rangle}{\sqrt{\delta{p_{a,f}}/\delta t}},\label{qt4}
\end{equation}

\begin{figure}[htbp]
    \centering
    \includegraphics[scale=0.55]{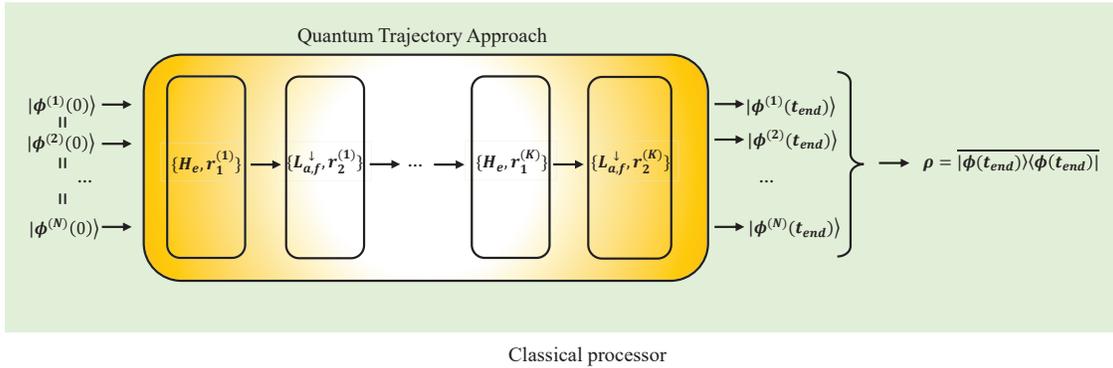}
    \caption{Detailed process of quantum trajectory approach boosting QAOA.}
    \label{qt}
\end{figure}

Normally, two random numbers $r_1$ and $r_2$, both between $0$ and $1$, are utilized to realize the probabilistic selection of $H_e$ and $L_{a,f}$. 
First, $r_1$ is generated.
If $r_1 > \delta p$, then no jump occurs, and the propagation follows Eq.~(\ref{qt3}). If $r_1 \le \delta p$, then a
jump occurs, and we must choose the particular jump operators to apply. Then, $r_2$ is generated to choose the smallest number of the jump operators dominating the evolution, which is $\{\tilde{Q}|\sum_{f=1}^{\tilde{Q}}\delta p^{\ \downarrow}_{a.f}\ge r_2\delta p\}$, and the propagation follows Eq.~(\ref{qt4}) with 
$L_{\tilde{Q},f}$.
Once $|\phi(t+\delta t)\rangle$ is obtained, $|\phi(t_{end})\rangle=|\phi(t+K\delta t)\rangle$ is at hand. Finally, average is performed over a large number of trajectories to ensure precision. Fig.~\ref{qt} depicts the detailed process of this procedure. 

The effectiveness of the quantum trajectory approach for the augmented system model can somewhat be verified through the first order expansion,
\begin{align}
    \overline{\rho(t+\delta t)}\,=\,& (1-\delta p)\frac{(1-i{ H}_e\delta t)|\phi(t)\rangle}{\sqrt{1-\delta_p}} \frac{\langle\phi(t)|(1+i{ H}_e^\dagger\delta t)}{\sqrt{1-\delta_p}}
    + \sum_{f=1}^Q \delta {p_{a,f}} \frac{\sqrt{\gamma_{_f}}L_{a,f}|\phi(t)\rangle}{\sqrt{\delta p_{a,f}/{\delta_t}}}
    \frac{\langle\phi(t)|L^{\dagger}_{a,f}\sqrt{\gamma_{_f}}}{\sqrt{\delta p_{a,f}/{\delta_t}}}\nonumber\\
    =\,& \rho(t)-i({ H}_e \rho(t)-\rho(t){ H}_e^\dagger)\delta t + \sum_{f=1}^Q \gamma_{_f} L_{a,f}\rho(t)L_{a,f}^\dagger\delta t,
    \label{qt_dm}
\end{align}
which is consistent to Eq.~(\ref{qt1}). Compared with Alg.~\ref{alg:pro}, the major difference of Alg.~\ref{alg:m} is the evolution process. For each trajectory, two lists of random numbers alternately decide the evolution of state vectors. After sufficient number of stochastic trajectories have been generated, an average of the final state vector is calculated to compute the final density matrix. With similar piecewise control of Hamiltonians of the principal system, one evolution process is completed. The whole procedure is detailed in Alg.~\ref{alg:m}. Note that the function $StochasticAverage()$ means that the evolution is determined by random numbers and the final density matrix is obtained via averaging over multiple trajectories.
\begin{algorithm}[htbp]
    \caption{Boosted and regularized QAOA in non-Markovian quantum systems}%标题
    \begin{algorithmic}[1]
      %这里是伪代码内容
     \Require an initial duration vector $\tau_1$, a learning rate $\upsilon$, a weight $\xi$, two Hamiltonians $H$ and $H'$,
      \Ensure the updated duration vector $\tau_k$, the state of the principal system $\rho_p(\tau_k)$ and its corresponding objective value $h(\tau_k)$,
      \State $k = 1$
      \State EVOLUTIONQT($\tau_k$, $H$, $H'$)
      \Do
      \State $k \leftarrow k+1$
      \State $d_k = \tau_{k-1}-\upsilon \nabla tr(H\rho_p(\tau_{k-1}))$
      \State $\tau_{k}=S_{\xi \upsilon}(d_k)$\Comment{the soft-thresholding operation}
      \State EVOLUTIONQT($\tau_k$, $H$, $H'$)
      \doWhile{$|h(\tau_k)-h(\tau_{k-1})|<\eta$}\Comment{the terminating condition}\\
      \Return $\tau_k,\rho_p(\tau_k), h(\tau_k)$\\
      \Function{Evolutionqt}{$\tau_k$, $H$, $H'$}
      \State $p=\frac{1}{2}|\tau_k|,|\phi(0)\rangle=|\phi_p(0)\rangle\otimes|\phi_a(0)\rangle$\Comment{$|\cdot|$ denotes the vector cardinality}
      \For{$j=1,\cdots, p$}
        \State  $H_p = H$ for $\beta_{jk}$ duration in Eq.~(\ref{qt2}) $\rightarrow$ $|\phi(t)\rangle$ $\rightarrow$ $|\phi(t_{end})\rangle$\Comment{one $|\phi(t_{end})\rangle$ for one trajectory}
        \State $\rho=StochasticAverage(|\phi(t_{end})\rangle\langle\phi(t_{end})|)$
        \State $H_p = H'$ for $\zeta_{jk}$ duration in Eq.~(\ref{qt2}) $\rightarrow$ $|\phi'(t)\rangle$ $\rightarrow$ $|\phi'(t_{end})\rangle$
        \State $\rho=StochasticAverage(|\phi'(t_{end})\rangle\langle\phi'(t_{end})|)$
      \EndFor 
      \State $\rho_p(\tau_k)$ $=$ ${\rm tr}_a[\rho]$\Comment{trace out the ancillary system}
      \State $h(\tau_k)={\rm tr}(H \rho_p(\tau_k))$
      \State \Return $h(\tau_k)$, $\rho_p(\tau_k)$
      \EndFunction
    \end{algorithmic}\label{alg:m}
\end{algorithm}
\section{Numerical Simulation of QAOA in a Non-Markovian Quantum System}\label{sec:sim}

To evaluate the performance of our algorithm, we will use QAOA to solve Max-Cut problem in a non-Markovian quantum system with different noises. In this section, we first brief the Max-Cut problem and corresponding setup for the simulation. Further, we give our solutions to the Max-Cut problem in a non-Markovian quantum system disturbed by quantum Lorentzian noise and explore how non-Markovianity affects the performance of our algorithm. In the end, larger scale  Max-Cut problem and QAOA in a non-Markovian quantum system disturbed by double-Lorentzian noise are studied for generality of our algorithm.

\subsection{Max-Cut and Approximation Ratio}\label{41}
% 图1 Max-Cut图
The performance of QAOA in non-Markovian quantum systems is demonstrated on a Max-Cut problem. This problem is a classical combinatorial optimization problem and is known to be NP-complete. The problem considers an $N$-node undirected yet weighted graph $G=(V,E)$, where $V$ is the node set and $E$ is the edge set. Max-Cut is the partition of $V$ into two subsets $V_1$ and $V_2$, where the aggregation of weights of crossing edges is maximized. If we assign $-1$ to vertices in $V_1$ and $1$ to vertices in $V_2$, Max-Cut can be formulated as a binary optimization problem, vice versa; i.e., $C = \max\sum_{(i,j)\in E}\frac{\omega_{ij}}{2}(1-s_i s_j),\  s_i,s_j \in \{-1,+1\}$ with the weight $\omega_{ij}$ for the $i$th and $j$th nodes.
For convenience, the equivalent form is used in this paper
$
    C = \min\sum_{(i,j)\in E}\omega_{ij}s_i s_j, \ s_i,s_j \in \{-1,+1\}
$, as the total weight is a constant.
The corresponding Hamiltonian obtained from Ising formulation is
\begin{equation}
    H = \sum_{(i,j)\in E}\omega_{ij}\sigma_{i}^z\sigma_{j}^z,
\end{equation}
and we want to minimize the expectation of $H$ as aforementioned.
We use the approximation ratio
\begin{equation}
    r = \frac{C_{max}-tr(H \rho_p(\hat{\tau}))}{C_{max}-C_{min}}\in [0,1],
\end{equation}
as a measure of how close the final state is to the optimal solution, where $C_{max}$ and $C_{min}$ are the theoretical maximum and minimum values of the original objective function, and $\rho_p(\hat\tau)$ is the obtained final density matrix by the algorithm. Obviously, a larger $r$ indicates a better solution represented by the final state. 

\subsection{Experimental Setting}\label{42}
\begin{figure}[htbp]
\centering
\includegraphics[scale=0.45]{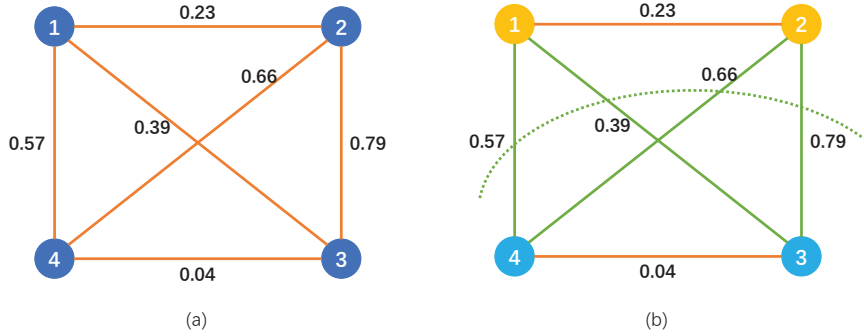}
\vspace{-2mm}
\caption{(a) is the undirected and weighted graph which is randomly generated for evaluating the performance of the proposed QAOA. The regular graph composes of $4$ nodes and $6$ differently-weighted edges. (b) is the optimal solution of our Max-Cut instance. The $4$ green edges are the crossing edges, and the dotted green curve is the Max-Cut (solution), which partitions the nodes into two categories (yellow and blue).}
\label{pic:weighted}
\end{figure}

We randomly generate an undirected yet weighted graph as depicted in Fig.~\ref{pic:weighted}(a). In this graph, $C_{max}=0.23+0.57+0.39+0.66+0.79+0.04=2.68$, where all the vertices are in the same group, while $C_{min}=-(0.57+0.39+0.66+0.79)+0.23+0.04=-2.14$, where the four vertices are partitioned into $\{1,2\}$ and $\{3,4\}$.

Further, we specify the spectrum of the non-Markovian environment as a Lorentzian spectrum
\begin{equation}
    S(\omega)=S_1(\omega) = \frac{\kappa_1\frac{\gamma_1^2}{4}}{\frac{\gamma_1^2}{4}+(\omega-\omega_a)^2},
    \label{eq:spec}
\end{equation}
where $\omega_a$ and $\gamma_1$ determine the center and the width of the spectrum. This environment is represented by a one-mode oscillator in the augmented system model, where $\omega_a$ and $\gamma_1$ also are the angular frequency and the damping rate to quantum white noise of the oscillator, respectively~\cite{article}. In addition, the coupling strength $\kappa_1$ determines the amplitude of the spectrum~\cite{article}. The dimension of the ancillary system is truncated to be $8\times 8$. In our instance, the principal system is comprised of $4$ qubits and its dimension of the density matrix is $16\times 16$. The Hamiltonian of the ancillary system is $H_a=\omega_aa^\dagger a$, where the angular frequency of the ancillary system is $\omega_a=10{\rm GHz}$. The coupling operator $L_{a,1}=a$, and $c_a=-\frac{\sqrt{\gamma_1}}{2}a$, where aforementioned $\gamma_1=0.6{\rm GHz}$. Each qubit of the principal system is coupled with the ancillary system through the direct coupling operator $z=\sqrt{\kappa_1}\sigma_y$, with the coupling strength $\kappa_1=1{\rm GHz}$. A Markovian system for QAOA is approximately generated with $\gamma_1\rightarrow+\infty$ when the Lorentzian noise reduces into the white noise.

\begin{figure}[ht]
\centering
\includegraphics[width=1.05\linewidth,height=80mm]{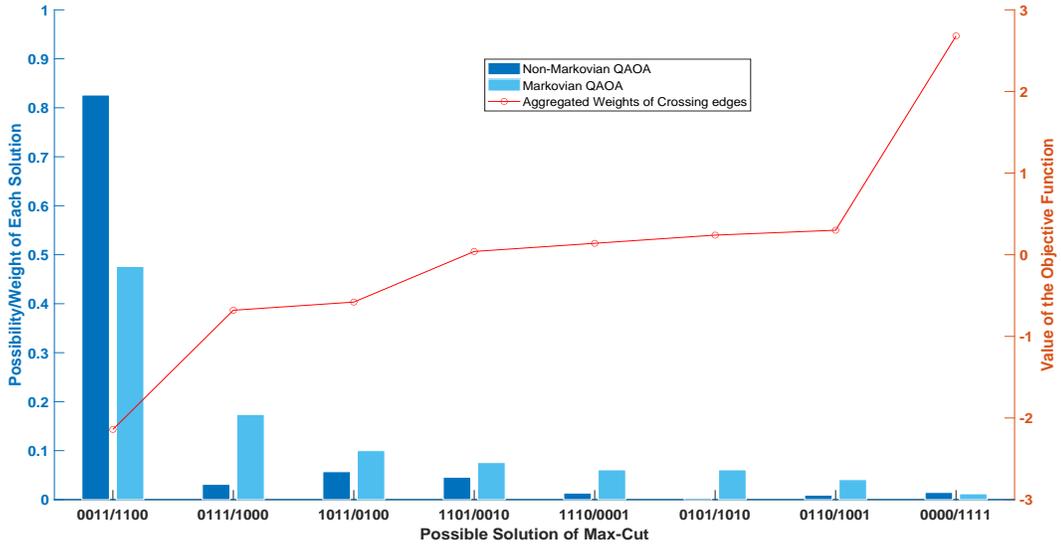}% Here is how to import EPS art
\caption{The possibility and value of each of the $8$ possible solutions in non-Markovian and Markovian QAOA. The bar chart follows the left blue $y$ axis, while line chart follows the right red $y$ axis. The left $y$ axis shows the possibility or weight of each solution in the final superpositioned quantum state, while the right $y$ axis shows the corresponding objective value for each solution, which is obviously the same for both non-Markovian and Markovian QAOA. The dark blue bar chart depicts the possibility of each solution in non-Markovian QAOA, while the light blue bar chart depicts that in Markovian QAOA. Both non-Markovian QAOA and Markovian QAOA are optimized based on initial depth $p=2$ and all control parameters initialized as $3$. As for the optimal solution, the weight in non-Markovian QAOA is 0.8260 given $(\zeta_1,\beta_1,\zeta_2,\beta_2)=(2.1,0.5,2.1,1.9)$, while the weight in Markovian QAOA is 0.4760 given $(\zeta_1,\beta_1,\zeta_2,\beta_2)=(0.6,2.9,1.2,1.0)$. Clearly, non-Markovian QAOA performs much better than Markovian QAOA. }\label{fig:possi}
\end{figure}

Fig.~\ref{fig:possi} shows the comparison of our Max-Cut instance in both non-Markovian and Markovian quantum systems. In our notation, $|0\rangle$ and $|1\rangle$ are used to specify the group which each of the four vertices belongs to. For example, $|0011\rangle$ means the vertices $1$ and $2$ are in the same group labeled by $|0\rangle$, while the vertices $3$ and $4$ are in the other group $|1\rangle$. Remarkably, $|1100\rangle$ outputs the same result as $|0011\rangle$, since the partition is the same in essence. As mentioned above, QAOA raises the possibilities of good solutions, and tends to choose these potentially optimal solutions when measurement is taken in quantum devices. We illustrate the possibility of each possible solution in Fig.~\ref{fig:possi}. The optimal solution $|1100\rangle$ or $|0011\rangle$ accounts for $0.8260$ in $1$, which is the largest among eight possible solutions. The corresponding value of the optimal solution is $-2.14$, which equals to $C_{min}$. Although both Markovian  and non-Markovian QAOA give the maximum possibility to the optimal solution in our experiment shown in Fig.\ref{pic:weighted}(b), non-Markovian QAOA has a higher probability for good solutions than Markovian QAOA, as shown in Fig.~\ref{fig:possi}.

\subsection{How Non-Markovianity Affects the Performance of QAOA}\label{43}

Next, we investigate how non-Markovianity affects the performance of QAOA. The definition of the measure of non-Markovianity in any quantum process is given and then we discuss the detailed impact of non-Markovianity on the performance of QAOA.
\subsubsection{Measure for the degree of non-Markovianity}
\textcolor{white}{1}\\
In this paper, we adopt the measure for the degree of non-Markovian behavior in open quantum systems constructed in
Ref.~\cite{Breuer2009MeasureFT}. The essential property of non-Markovianity is the growth of distinguishability between two quantum states under a quantum evolution, which can be interpreted as a flow of information from the environment back to the open system. The measure for non-Markovianity in Ref.~\cite{Breuer2009MeasureFT} is based on the trace distance of two quantum states
\begin{equation}
    D(\rho_{_1},\rho_{_2})=\frac{1}{2}{\rm tr}|\rho_{_1}-\rho_{_2}|,
\end{equation}
describing the probability of successfully distinguishing the two states,
where $\rho_{_1}$ and $\rho_{_2}$ are two time-evolving density matrices of %\textcolor{red}{
the same pre-defined quantum processes %\textcolor{red}{
for different inputs%}
, and $|A|=\sqrt{A^\dagger A}$ for an arbitrary operator $A$. In a time sequence $t_0, t_1, \cdots, t_L$ with appropriate intervals, the degree of non-Markovianity is calculated as
\begin{align}
     \mathcal{N}(\Phi)=&\ \max_{\rho_{_{1,2}}}\sum_{i=0}^{L-1} \frac{1+sgn(\Delta D_i)}{2} \Delta D_i,
\end{align}
where $\Delta D_i =  \ D(\rho_{_1}(t_{i+1}),\rho_{_2}(t_{i+1}))-D(\rho_{_1}(t_i),\rho_{_2}(t_i))$ and $sgn()$ is the sign function that is used to only accepts the increase in $D$.
%% 和辅助系统维数无关
\subsubsection{Non-Markovianity and performance of QAOA}
\textcolor{white}{1}\\
In addition, in order to calculate the non-Markovianity of the system for QAOA, we choose the state of the principal system as $\rho_{_1}(0)={|+\rangle}^4{\langle+|}^4$ and $\rho_{_2}(0)={|-\rangle}^4{\langle-|}^4$%$\textcolor{red}{n=?} 
to maximize the sum of the growth of $D$ as suggested in Ref.~\cite{Breuer2009MeasureFT}. Hence, in the following we will observe how these parameters vary the non-Markovianity $\mathcal{N}(\Phi)$ of the system of interest and the relation between the non-Markovianity and the performance of QAOA. The optimized control durations are obtained via Alg.~\ref{alg:m}.
\begin{figure}[htbp]
\hspace{-33pt}
\begin{minipage}[H]
{0.6\linewidth}
\includegraphics[width=\linewidth]{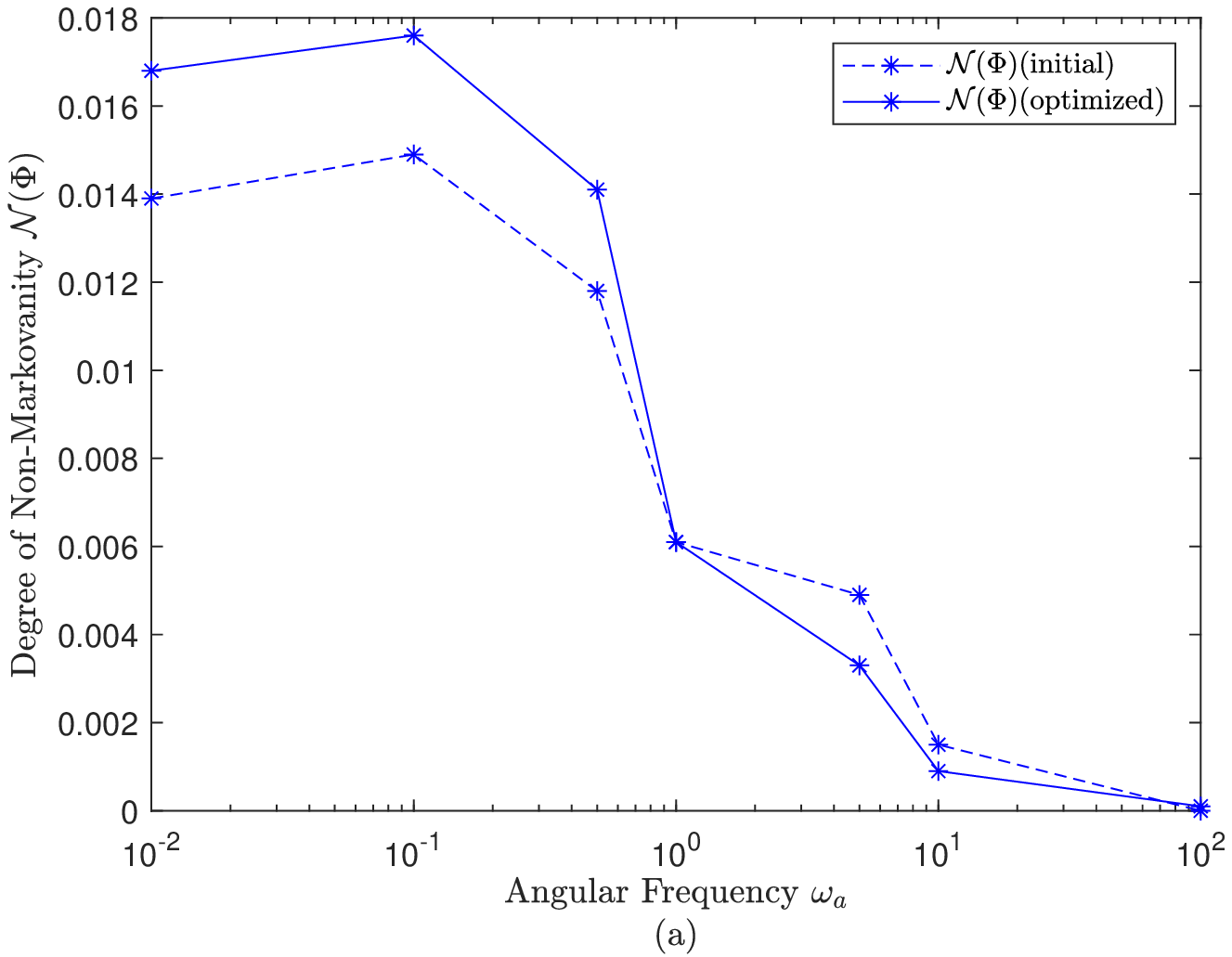}
\end{minipage}
\begin{minipage}[H]
{0.6\linewidth}
\includegraphics[width=\linewidth]{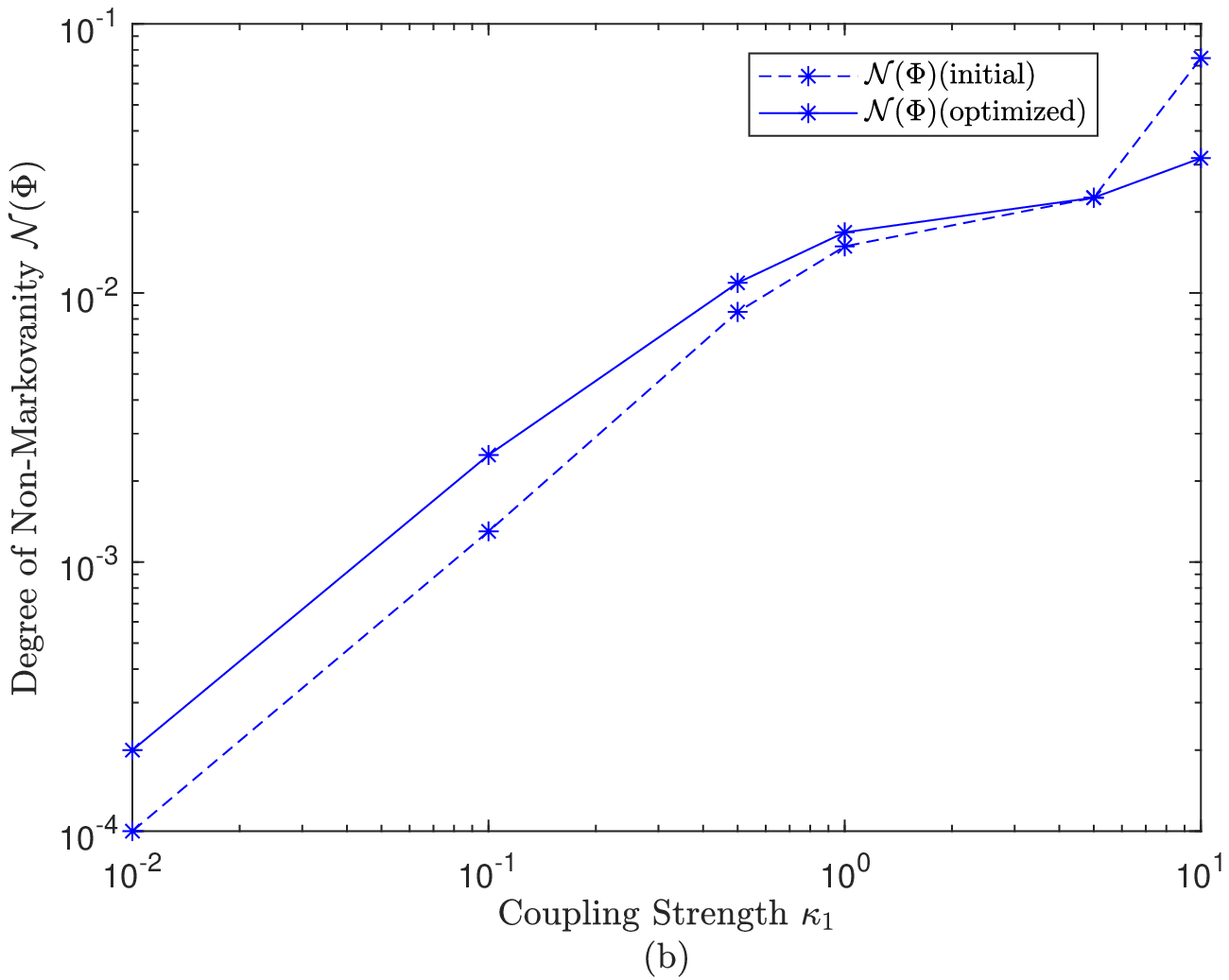}
\end{minipage}
\begin{minipage}[H]
{0.6\linewidth}
\hspace{-33pt}
\includegraphics[width=\linewidth]{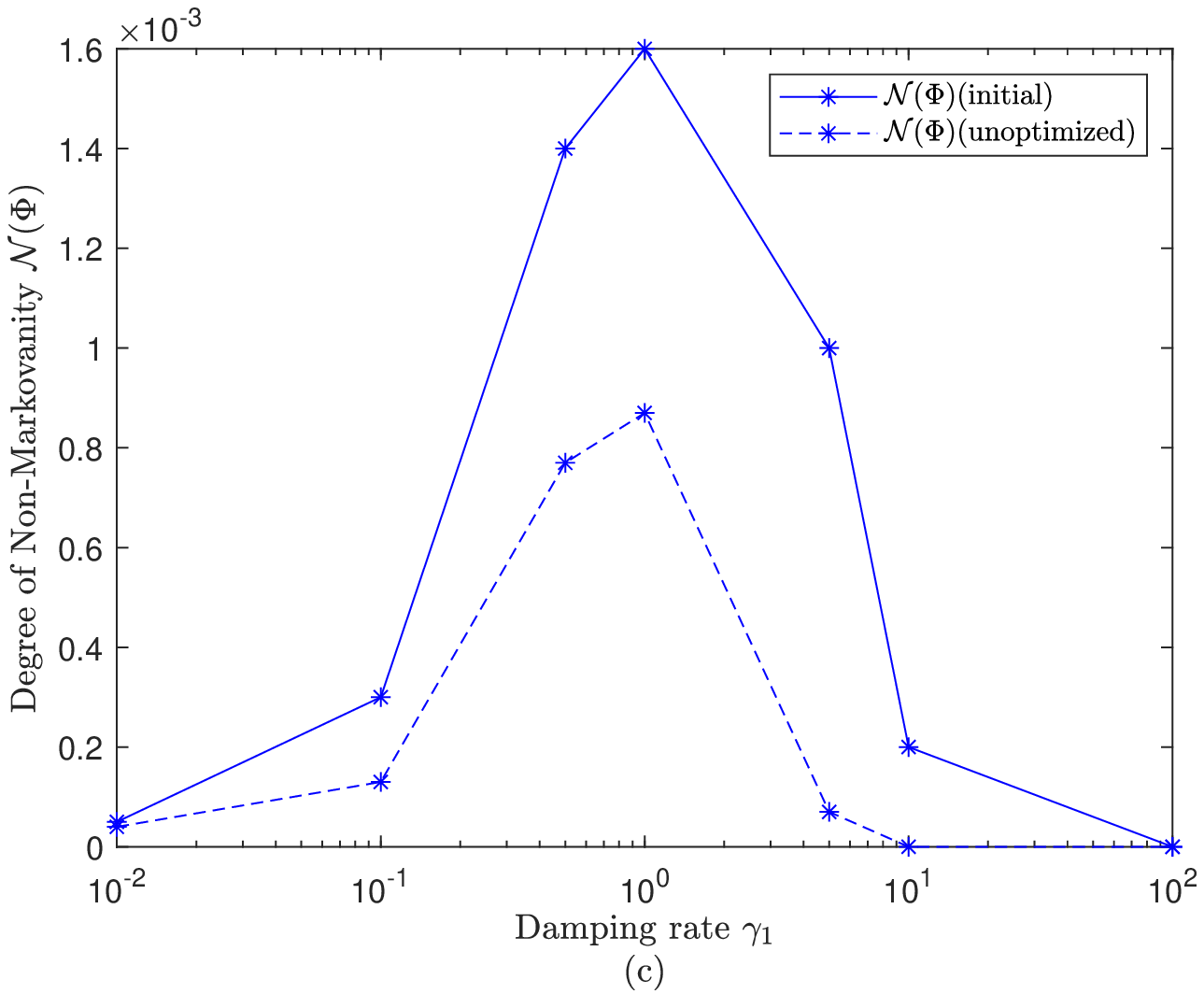}
\end{minipage}
\begin{minipage}[H]
{0.6\linewidth}
\hspace{-40pt}
\vspace{-11pt}
\includegraphics[width=\linewidth]{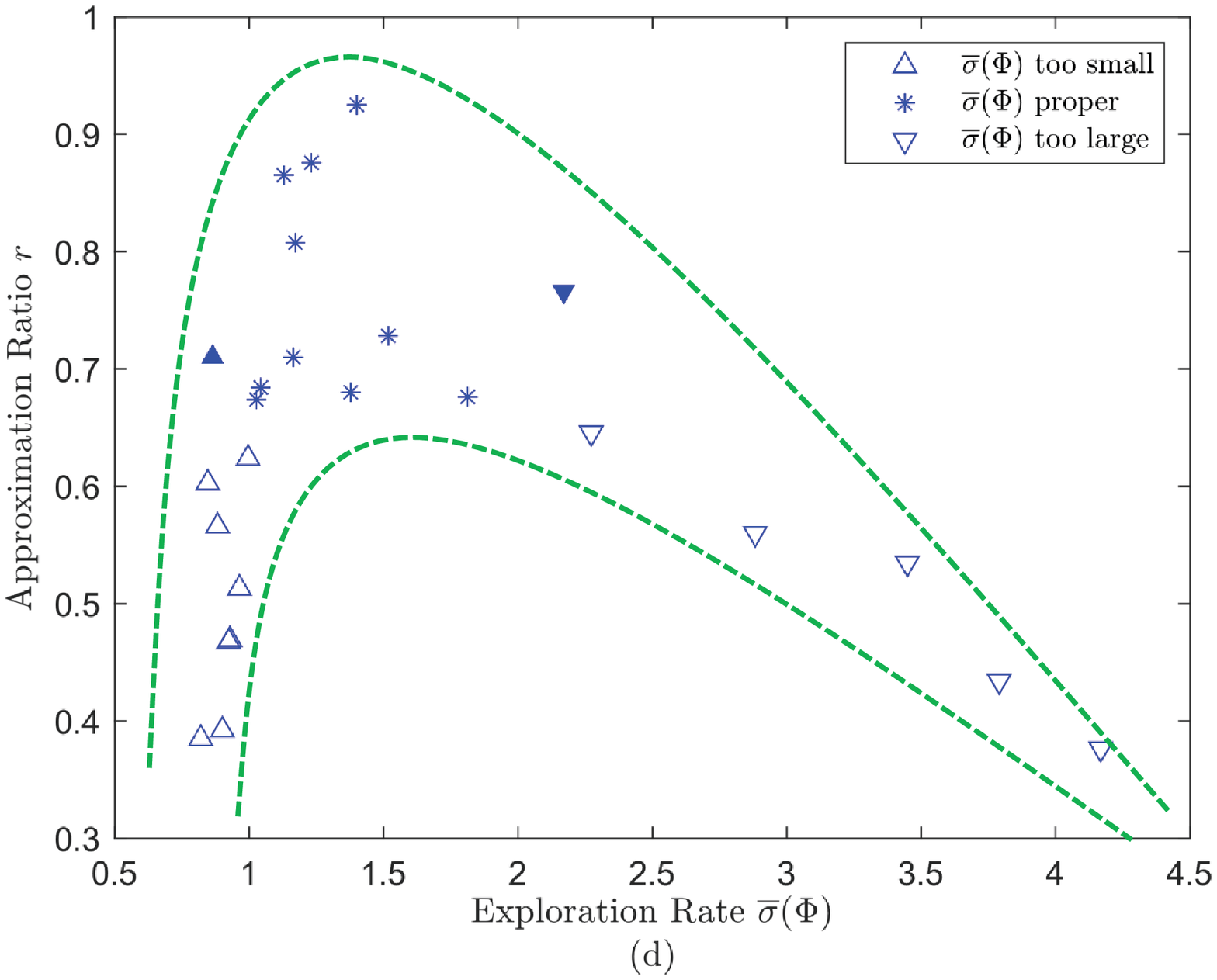}
\end{minipage}
\caption{The influence of angular frequency $\omega_a$, coupling strength $\kappa_1$ and damping rate $\gamma_1$ on the degree of non-Markovianity $\mathcal{N}(\Phi)$, and the influence of exploration rate $\overline{\sigma}(\Phi)$ on the approximation ratio $r$. For (a)(b)(c), the dashed line is obtained by initial control parameters, which is initialized as $3$, and $P=2$, while the solid line is obtained by optimized control parameters which varies at each point. For (d), the experiment is conducted under the condition that $P=2$, and each element of $\tau$ is randomly selected between $0.5$ and $4$. Since $\overline{\sigma}(\Phi)$ can only be measured after the evolution of a quantum process, scatter plot is applied to demonstrate our results.}\label{par}
\end{figure}

Fig.~\ref{par}(a), (b), and (c) depict the influence of $\omega_a$, $\kappa_1$ and $\gamma_1$ on $\mathcal{N}(\Phi)$. 
From Fig.~\ref{par}(a), we find that %non-Markovianity
$\mathcal{N}(\Phi)$ reaches its peak when the principal system resonates with the ancillary system. In this case, the principal system is significantly disturbed by the ancillary system. In Fig.~\ref{par}(b), $\mathcal{N}(\Phi)$ increases monotonically with $\kappa_1$. This is because a large $\kappa_1$ means the coupling strength between the principal system and the ancillary system is large, which leads a large $\mathcal{N}(\Phi)$. In Fig.~\ref{par}(c), $\mathcal{N}(\Phi)$ first increases with $\gamma_1$, and reaches its peak at $\gamma_1=1{\rm GHz}$, and then decreases with $\gamma_1$. This is due to the Lorentzian power spectral density~(\ref{eq:spec}),
where $\omega_a$ is the angular frequency of the ancillary system. It is clear that when $\gamma_1$ approaches $0$ or $+\infty$, the Lorentzian noise reduces into the white noise, and the principal system turns into a Markovian system, causing $\mathcal{N}(\Phi)$ to approach $0$.
Additionally, Fig.~\ref{par}(a), (b), and (c) indicate that the increase in the performance of QAOA, namely from the initial to the optimized is accompanied by the increase in the degree of non-Markovianity at certain points. This deserves further discussion. %\textcolor{red}{
Note that in the three figures, all the data points are discrete, since one parameter setting (angular frequency, coupling strength and damping rate) produces one measured degree of non-Markovianity. The polylines are connection of adjacent data points with regard to angular frequency, coupling strength or damping rate. This is intended to show how the degree of non-Markovianity varies with regard to them. The three x-axes are of logarithmic coordinates, and we choose these data points to show the trend of the degree of non-Markovianity on a large scale.%} 

As aforementioned, the essential property of the non-Markovian behavior is the growth of distinguishability between quantum states. A large value of  $\mathcal{N}(\Phi)$ indicates the fact that the density matrix tends to vary drastically, thus exploring the potential solutions represented by quantum states more boldly. This is the situation for basic case. Nonetheless, if  $\mathcal{N}(\Phi)$ exceeds a certain amount, the exploration process becomes too fast to converge to a good enough solution, leading to a low $r$. This resembles the idea ``balance between exploration and exploitation" in Reinforcement Learning~\cite{coggan2004exploration}. Only if both exploration and exploitation are properly taken into account can we obtain a good result. In our paper, the evolution of QAOA is responsible for exploration and PG algorithm accounts for exploitation. Thus, a proper exploration procedure is preferred.

In order to better characterize the significance of exploration on the approximation ratio $r$, we dub
\begin{equation}\label{Exrate}
 \overline{\sigma}(\Phi)=\frac{\mathcal{N}(\Phi)}{T_e}=\frac{\mathcal{N}(\Phi)}{\sum_{i=0}^{L-1}\frac{1+sgn(\Delta D_i)}{2}(t_{i+1}-t_i)}=\frac{2\mathcal{N}(\Phi)}{\sum_{i=0}^{L-1}(1+sgn(\Delta D_i))(t_{i+1}-t_i)}
\end{equation}
the exploration rate, which signifies the average degree of exploring quantum states in a quantum process $\Phi(t)$, and $T_e$ denotes the total amount of time when $\mathcal{N}(\Phi)$ increases within $\Phi(t)$. The exploration rate reflects the speed of growth in the degree of non-Markovianity, which is independent of time. Hence, it can serve as an indicator of how fast the density matrix explores in the metric space where distance denotes the half difference in trace between the two density matrices. We randomly select $28$ groups of control durations without optimization, and obtain corresponding $\overline{\sigma}(\Phi)$ and $r$, depicted in Fig.~\ref{par}(d). Clearly, when $\overline{\sigma}(\Phi)$ is smaller than $1$, the exploration is far from enough that a good solution is hardly detected. When $\overline{\sigma}(\Phi)$ is between $1$ and $2$, we consider the exploration of good solutions is relatively sufficient. However, when $\overline{\sigma}(\Phi)$ continues to grow above $2$, the step for exploration becomes so large that good solutions would be skipped. The filled upside triangle denotes a quantum process that is more sensitive to $\overline{\sigma}(\Phi)$, and can find a relatively good solution even when $\overline{\sigma}(\Phi)$ is not large enough. On the contrary, the filled downside triangle denotes a quantum process that is not sensitive to $\overline{\sigma}(\Phi)$, and is able to find a relatively good solution even when $\overline{\sigma}(\Phi)$ is large. We use green dashed lines to denote the upper bound and lower bound of these data points. So we can clearly see a trade-off for exploration rate here.

To sum up, %\textcolor{red}{
non-Markovianity can be utilized as a quantum resource to help achieve a relatively good performance of QAOA in non-Markovian systems. This is characterized by a proper degree of exploration rate %\textcolor{red}{
under BLP-measure, which can serve as a guidance when optimizing control parameters. %\textcolor{red}{
Similar results can be obtained when a different measure of non-Markovianity, e.g., divisibility criteria, is applied. This is because different measures are all determined by the parameters of the augmented systems which affect the performance of QAOA.%}

\subsection{Testing of the Boosted Algorithm with Application to Complicated Graph Cases}\label{44}
For the augmented system, where $N_p$ and $N_a$ denote the dimensions of $\mathcal{H}_p$ and $\mathcal{H}_a$, Alg.~\ref{alg:m} only calculates the state vector of dimension $N_p \cdot N_a$, instead of the density matrix of dimension $(N_p \cdot N_a) \times (N_p \cdot N_a)$. %\textcolor{red}{
Although multiple runs have to be performed,%}
this feature avoids dimension explosion, %\textcolor{red}{
which is especially beneficial when dealing with large open quantum systems. In addition, parallel computing can be applied to improve computation efficiency, since trajectories do not rely on one another. To demonstrate the efficiency of Alg.~\ref{alg:m} versus Alg.~\ref{alg:pro}, we conduct numerical tests under fair conditions. Table~\ref{eff} shows the comparison of computation efficiency between Alg.~\ref{alg:m} and Alg.~\ref{alg:pro}, where $T$ and $M$ denotes time and memory overhead, while $500$ and  $1000$ are the number of trajectories averaged in the end. Apparently, as the number of nodes increases, Alg.~\ref{alg:m} becomes more efficient, which spends less computational time and memory overhead.
When the number of Max-Cut nodes overtakes $9$, Alg.~\ref{alg:pro} fails to run due to lack of preoccupied memory, while Alg.~\ref{alg:m} still works. %Clearly, as the number of Max-Cut nodes increases, Alg.~\ref{alg:m} has increasingly higher computation efficiency. 

\begin{table}[htb]
  \centering
  \caption{Computation efficiency of Alg.~\ref{alg:m} versus Alg.~\ref{alg:pro}}
  \label{tab1}
  \begin{tabular}{c|cccccc}
    \br
   Number of Max-Cut nodes & 5 & 6 & 7 & 8 & 9\\ \hline
    $T_{Alg2(500)}/T_{Alg1}$  & 0.422 & 0.192& 0.144 & 0.103 & Alg1-MemoryError\\ \hline
    $M_{Alg2(500)}/M_{Alg1}$ & 0.235 & 0.138& 0.096 & 0.061 & Alg1-MemoryError\\ 
    \hline
    $T_{Alg2(1000)}/T_{Alg1}$ & 0.664 & 0.232& 0.162 & 0.112 & Alg1-MemoryError\\ \hline
    $M_{Alg2(1000)}/M_{Alg1}$ & 0.387 &0.150 & 0.117 & 0.087 & Alg1-MemoryError\\ 
    \br
  \end{tabular}\label{eff}
\end{table}

\begin{figure}[H]
\centering
\includegraphics[width=0.8\linewidth]{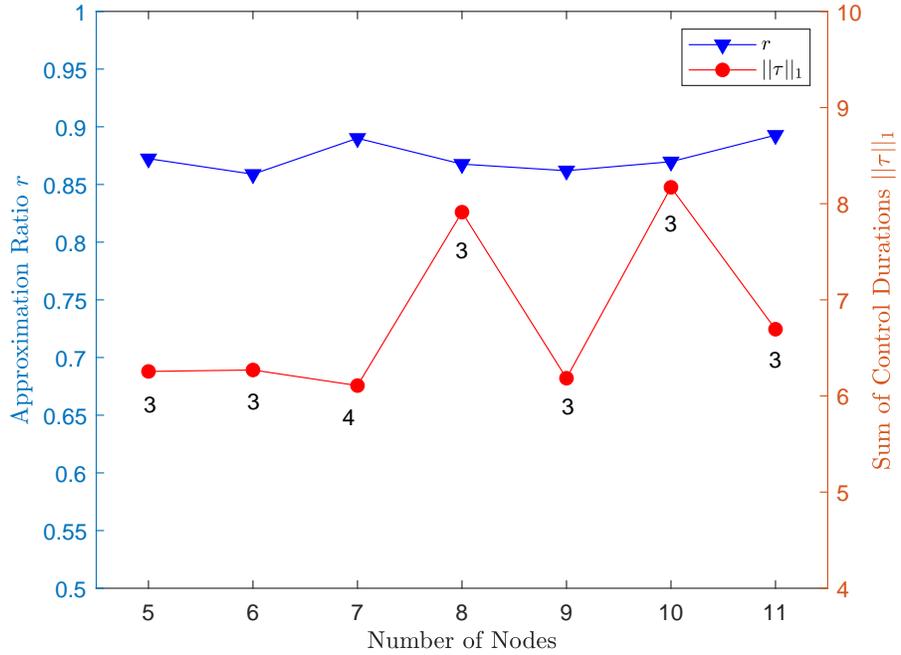}
\caption{The approximation ratio $r$ and sum of control durations $||\tau||_1$ versus number of nodes from $5$ to $11$. The left blue $y$ axis shows the approximation ratio for each Max-Cut case, while the right red $y$ axis indicates the sum of all control durations, namely $||\tau||_1$. The number below the red data point denotes the optimized control depth $P$. The initial values of control parameters are randomly chosen between $0.5$ and $4$.}
    \label{massive}
\end{figure}

For generality, we conduct diverse Max-Cut cases with weighted graph containing five nodes to eleven nodes. We randomly generate weights for edges in Table~\ref{div}. Corresponding edges with weights are selected, according to the number of nodes. Specifically, if the number of nodes is set as $7$, then all the edges  without repetition that include nodes from $Z_0$ to $Z_6$ are selected with corresponding weights. Through Alg.~\ref{alg:m}, optimized approximation ratio and depth are shown in Fig.~\ref{massive}. 
%\textcolor{red}{
In this figure, all the data points are discrete, since each graph, denoted by the number of nodes, produces an approximation ratio r and control duration ${|\left|\tau\right||}_1$. The blue polyline is connection of adjacent data points with regard to number of nodes. The blue polyline shows the approximation ratio is high and stable, between $0.85$ and $0.9$, when the graph becomes more sophisticated. Likewise, the red polyline is connection of adjacent date points with regard to number of nodes. The red polyline demonstrates how the sum of control duration varies when the graph becomes more complex, which manifests that the sum of control duration reduces to a small value despite a large initial value.
%}
%The approximation ratios are all above $0.85$.
The two consistencies verify the effectiveness and generality of Alg.~\ref{alg:m}.

\begin{table}[ht]
    \centering
    \caption{Randomized weights for five-node to eleven-node Max-Cut cases.}
    \scalebox{1}{
    \begin{tabular}{c|ccccccccccc}
\br
&$Z_0$&$Z_1$&$Z_2$&$Z_3$&$Z_4$&$Z_5$&$Z_6$&$Z_7$&$Z_8$&$Z_9$&$Z_{10}$\\
\hline
$Z_0$&0&0.60&0.79&0.71&0.40&0.66&0.33&0.50&0.27&0.88&0.47\\
$Z_1$&*&0&0.03&0.21&0.56&0.72&0.82&0.81&0.66&0.73&0.53\\
$Z_2$&*&*&0&0.65&0.75&0.46&0.86&0.38&0.66&0.32&0.85\\
$Z_3$&*&*&*&0&0.54&0.09&0.77&0.96&0.99&0.35&0.66\\
$Z_4$&*&*&*&*&0&0.41&0.16&0.79&0.56&0.63&0.85\\
$Z_5$&*&*&*&*&*&0&0.22&0.36&0.34&0.33&0.44\\
$Z_6$&*&*&*&*&*&*&0&0.76&0.01&0.62&0.42\\
$Z_7$&*&*&*&*&*&*&*&0&0.57&0.13&0.79\\
$Z_8$&*&*&*&*&*&*&*&*&0&0.93&0.17\\
$Z_9$&*&*&*&*&*&*&*&*&*&0&0.73\\
$Z_{10}$&*&*&*&*&*&*&*&*&*&*&0\\
\br
\end{tabular}}
\label{div}
\end{table}

\subsection{Case with Different Noise}\label{45}
Besides Lorentzian noise, quantum colored noise arising from a non-Markovian environment may have various power spectral
densities. Since arbitrary power spectral densities of quantum colored noise can be expressed as a combination of Lorentzian power spectral densities, the internal modes of
the non-Markovian environment can be represented by concatenated ancillary systems~\cite{multi}. Here, we give an example of a non-Markovian quantum system disturbed by double Lorentzian noise where two oscillators are coupled to the principal system. The Max-Cut scenario and the way to calculate the exploration rate here is the same as the previous four-node case. The parameters for the first oscillator are kept as those in our previous simulation and the parameters for the second oscillator are given as $\omega_{a2}=5{\rm GHz}$, $\gamma_2=1{\rm GHz}$ and $\kappa_2=0.8{\rm GHz}$. Thus the total resulting power spectrum density is
\begin{equation}
    S(\omega) = S_1(\omega) + S_2(\omega) = \frac{\kappa_1\frac{\gamma_1^2}{4}}{\frac{\gamma_1^2}{4}+(\omega-\omega_a)^2}+\frac{\kappa_2\frac{\gamma_2^2}{4}}{\frac{\gamma_2^2}{4}+(\omega-\omega_{a2})^2},
\end{equation}
and the corresponding master equation for the augmented system is written as
\begin{equation}
    \Dot{\rho}(t) = -i[H_p+H_a+H_{a2}+H_{pa}+H_{pa2}, \rho(t)] +\gamma_1\mathcal{L}^*_{L_{a,1}}(\rho(t)) + \gamma_2 \mathcal{L}^*_{L_{a,2}}(\rho(t)),
\end{equation}
where $H_{a2}=\omega_{a2}a_2^\dagger a_2$ is the Hamiltonian of the second ancillary system, and $H_{pa2}=i(c_{a2}^\dagger z_2-z_2^\dagger c_{a2})$ is the second direct interaction Hamiltonian between the principal system and the second ancillary system, both defined on the Hilbert space of the second ancillary system $\mathcal{H}_{a2}$. 
The fictitious output for the second ancillary system is $c_{a2}=-\frac{\sqrt{\gamma_2}}{2}a_2$, and the second direct coupling operator is $z_2=\sqrt{\kappa_2}\sigma_y$.

\begin{figure}[!ht]
\centering
\includegraphics[width=0.8\linewidth]{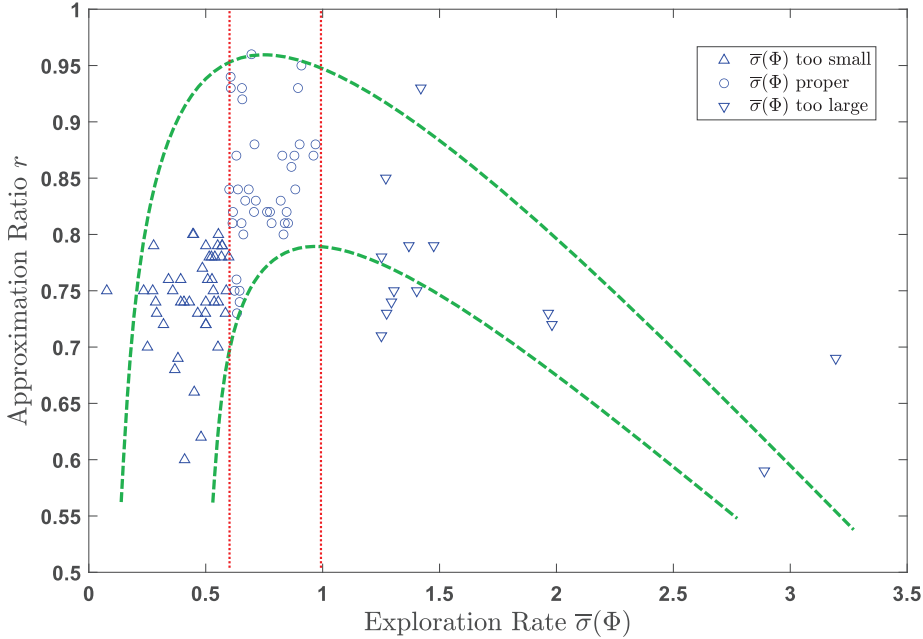}
\caption{The influence of exploration rate $\overline{\sigma}(\Phi)$ versus the approximation ratio $r$ on the case with double-Lorentzian noise. The experiment is conducted with $100$ data points under the condition that $P=2$, and each element of $\tau$ is also randomly selected. The upside triangles denote the data points with small $\overline{\sigma}(\Phi)$, the stars denote the data points with proper $\overline{\sigma}(\Phi)$, and the downside triangles denote the data points with large $\overline{\sigma}(\Phi)$. The two green dashed lines denote the trend of these data points.}\label{dif}
\end{figure}

This case achieves the same result as Fig.\ref{pic:weighted}(b) via Alg.~\ref{alg:m}, where maximum probability is assigned to the optimal solution $|0011\rangle$ and $|1100\rangle$.
Similar to Fig.~\ref{par}(d), we draw the approximation ratio varying with the exploration rate in Fig.~\ref{dif}, where a proper degree of $\overline{\sigma}(\Phi)$  in $0.6 
 <\overline{\sigma}(\Phi)<1$ is also in favor of good solutions. 
%\textcolor{red}{
In Fig.~\ref{dif}, the two trend lines indicate the upper bound and lower bound for ordinary data points. We show that a proper amount of exploration rate would benefit the approximation ratio most, which means a medium exploration rate achieves a higher approximation ratio than a small exploration rate or a large exploration rate.%}
When $\overline{\sigma}(\Phi)\le0.6$ the exploration process is relatively conservative and unlikely to access good solutions. When $\overline{\sigma}(\Phi)\ge 1$, the exploration process is drastic that good solutions may be skipped. Note that there are some exceptions, since the evolution and optimization procedure contain a certain degree of randomness, it is reasonable that good solutions are sometimes seized with $\overline{\sigma}(\Phi)$ beyond proper amount. However, a proper $\overline{\sigma}(\Phi)$ can serve as a necessary condition for good solutions for a majority of conditions. In summary, this example verifies our discovery that non-Markovianity can help with QAOA to obtain a good solution, which can be indicated by $\overline{\sigma}(\Phi)$. Also, the presented framework can be generalized to run QAOA in a non-Markovian quantum system with an arbitrary spectrum.

\section{Conclusion %\textcolor{red}{
and Future Work}%}
\label{sec:six}
In our work, we have presented QAOA in non-Markovian quantum systems with an augmented system model framework. Based on the augmented model, we have first proposed the regularized QAOA in non-Markovian systems. Then, in order to %\textcolor{red}{
alleviate the computational burden induced by the high dimensional augmented system, we have further proposed a boosted algorithm via quantum trajectory approach. The effectiveness of the above algorithm is verified in an example of the Max-Cut problem, where the algorithm runs in a non-Markovian quantum system disturbed by Lorentzian noise. Our algorithm can achieve a better performance than that in Markovian quantum systems and works for the Max-Cut problem with more nodes and under complicated noise. Remarkably, we find a proper exploration rate can help to obtain a better result, which should be balanced between exploration and exploitation of potential solutions. This framework works for non-Markovian quantum systems with an arbitrary spectrum such that it is potentially easy to work on NISQ devices, thus paving the way for efficiently addressing combinatorial optimization problems.

%\textcolor{red}{
There are several future research directions worth exploring. First, an intriguing direction is to control the degree of non-Markovianity such that the performance of QAOA on NISQ devices is enhanced. Further, it remains an open question how to utilize or allocate quantum decoherence as a quantum resource to boost indicators of system level instead of destroying or washing out important quantum effects.%}

%\Acknowledgements{This work is supported by the National Natural Science Foundation of China (NSFC) under Grants No. 62273226 and No. 61873162. This work is also supported by the Open Research Project of the State Key Laboratory of Industrial Control Technology, Zhejiang University, China (No. ICT2022B47).}

\end{document}